\documentclass[12pt, a4paper] {article}
\usepackage{amsfonts}
\usepackage{verbatim}
\usepackage{latexsym}
\usepackage{amsmath}

\begin{document}

\begin{center}
{\bf \small SUPERFIELD APPROACH TO NILPOTENT SYMMETRIES OF THE FREEDMAN-TOWNSEND MODEL: NOVEL FEATURES}

\vskip 1.5cm
{\small R. P. MALIK$^{(1,2,3)}$}\\
$^{(1)}$ {\it AS-ICTP, Strada Costiera, Trieste, Italy}\\
$^{(2)}$ {\it Physics Department, BHU, Varanasi, India}\\
$^{(3)}$ {\it DST-CIMS, BHU, Varanasi, India}\\

\vskip 0.05cm

{\small  {\bf e-mail: malik@bhu.ac.in}}
\end{center}

\vskip 2 cm

\noindent
{\bf Abstract:} We perform the Becchi-Rouet-Stora-Tyutin (BRST) analysis of the
Freedman-Townsend (FT) model of topologically massive non-Abelian theory by
exploiting its (1-form) Yang-Mills (YM) gauge transformations  to show
the existence of some {\it novel features} that are totally different from the results obtained
in such a kind of consideration carried out for the dynamical non-Abelian 2-form theory.
We tap here the potential and power of the ``augmented'' version of Bonora-Tonin's superfield approach
to BRST formalism to derive the full set of off-shell nilpotent and absolutely anticommuting
(anti-)BRST symmetry transformations where, in addition to the horizontality
condition (HC), we are theoretically compelled to exploit the 
appropriate gauge-invariant restrictions (GIRs) on the (super)fields
for the derivation of the appropriate symmetry transformations for {\it all} the relevant fields. 
We compare our key results with that
of the other such attempt for the discussion of the present model within the
framework of BRST formalism.\\



\newpage

\noindent
{\large \bf 1. Introduction}\\

\noindent
During the last few years, there has been a great deal of interest in the understanding of 
higher $p$-form ($p = 2, 3, 4,...$) gauge theories because of their relevance in the context
of modern developments in (super)string theories and related extended objects (see, e.g. [1-3]).
In particular, the  merging of 2-form and 1-form (non-)Abelian fields, through the celebrated
$B \wedge F$ term, has led to the development of topologically massive gauge models which provide
an alternative to the Higgs mechanism of standard model of particle physics. To be specific, the
(non-)Abelian 1-form gauge field acquires a mass in the topologically massive gauge theories (TMGTs)
without presence of any Higgs scalar fields [4-9]. In view of the fact that the Higgs particles have
not yet been observed experimentally, the above topologically massive models have generated a renewed
interest for their study and understanding.

In the context of the above TMGTs, it is pertinent to point out that we have studied the Abelian
version [10] of the TMGTs within the framework of BRST formalism and have also applied the superfield
approach to this system. One of the novel observations, in this realm of investigations,
has been the emergence and existence of the Curci-Ferrari (CF) type restriction even in the
case of this Abelian theory\footnote{It should be recalled that the CF condition [11] appeared, for the first time,
in the context of the BRST formulation of the 4D non-Abelian 1-form gauge theory.}. After applications of the geometrical
superfield approaches (see, e.g. [12-15]) to various $p$-form gauge theories, we have claimed
that one of the essential features of a gauge theory, described within the framework of  BRST
formalism, is the existence of the CF type restrictions. We have also shown their deep connections
with the geometrical object called gerbes in our systematic study of the (non-)Abelian 1-form
as well as the Abelian 2-form and 3-form gauge theories [16,17].

In particular, we have applied the Bonora-Tonin (BT) superfield approach [12,13] to the 4D dynamical
non-Abelian 2-form gauge theory (which happens to be a version of the non-Abelian TMGTs) and obtained the
off-shell nilpotent and absolutely anticommuting (anti-)BRST symmetry transformations corresponding
to its (1-form) Yang-Mills (YM) gauge symmetries as well as its (2-form) tensor gauge symmetries [18].
In a recent couple of papers [19,20], we have shown the existence of some novel features in the
BRST analysis of the above dynamical non-Abelian 2-form theory. To be specific, it has been shown
that the nilpotent and conserved (anti-)BRST charges, corresponding to the above YM symmetries as well as tensor
gauge symmetries, are {\it not} capable of generating the (anti-)BRST symmetries for some specific fields of
the theory. It has been also demonstrated, in these works [19,20], that even the requirements of
the nolpotency and absolute anticommutativity properties of the (anti-)BRST symmetry transformations,
are {\it not} able to generate the (anti-)BRST transformations of the above cited
specific fields.

In our present investigation, we concentrate on the study of  4D non-Abelian version of
TMGT, proposed by Freedman and Townsend (FT) [4], within the framework of BT superfield formalism
[12,13] to derive the off-shell nilpotent and absolutely anticommuting (anti-)BRST symmetry transformations
corresponding to {\it only} the usual (1-form) YM gauge symmetry transformations of this theory. We demonstrate
that the conserved  and nilpotent (anti-)BRST charges of our present analysis are {\it not} able
to generate the (anti-)BRST symmetry transformations for the 2-form field $B_{\mu\nu}$ (which happens
to be an auxiliary field in the FT model of non-Abelian TMGT). This is a novel observation
{\it vis-{\`a}-vis} our earlier study of the dynamical non-Abelian 2-form gauge theory [19,20].
We lay emphasis on the fact that it is the ``augmented''
version of the BT superfield formalism, of our present investigation, that is capable of obtaining the (anti-)BRST
symmetry transformations for the above auxiliary field ($B_{\mu\nu}$) but the standard generators
(i.e. (anti-)BRST charges) as well as the sacrosanct requirements of the nilpotency and
anticommutativity properties 
are {\it unable} to generate the above symmetries for the $B_{\mu\nu}$ field of the FT model (of
the non-Abelian version of TMGT).

The main motivating factors behind our present investigation are as follows. First and foremost,
the topologically massive 4D (non-)Abelian gauge theories are interesting in their own right as they
provide an alternative to the celebrated Higgs mechanism for the generation of mass of the
(non-)Abelian 1-form gauge fields. Second, it is very important to perform a comparative study of the
existing FT model [4] and dynamical 2-form [6,7,9] topologically massive non-Abelian gauge model so that
their relative strengths and weaknesses could become clear. Finally, the problem of constructing
a renormalizable, consistent and unitary non-Abelian 2-form gauge theory is still an 
open challenge.
Thus, any deeper understanding of the existing models [4,6], under any systematic scheme, is a welcome
step in the direction of achieving our central goal of constructing this theory in its full generality.

Our present paper is organized as follows. In Sec. 2, we briefly
recapitulate the bare essentials of the (1-form) YM  and (2-form) tensor
gauge symmetry transformations in the Lagrangian formulation. Our
Sec. 3 deals with the derivation of the proper (anti-)BRST
symmetry transformations associated with {\it all} the physical fields and
corresponding (anti-)ghost fields (along with the Curci-Ferrari condition)
within the framework of ``augmented'' BT
superfield approach to BRST formalism.  Sec. 4 and  Sec. 5
are devoted to the derivation of (anti-)BRST charges from the
(anti-)BRST symmetries of the coupled Lagrangian densities. The
conserved ghost charge and BRST algebra are deduced in Sec. 6.
Finally, we summarize our key results, comment on some subtle issues and make some
concluding remarks in Sec. 7.\\

\noindent
{\large \bf 2. Preliminaries: continuous gauge symmetries}\\

\noindent
Let us begin with the following four (3 + 1)-dimensional (4D) Lagrangian density,
proposed by Freedman-Townsend (hereafter called FT model), for the topologically massive
non-Abelian gauge theory\footnote{We
take the signature of the flat metric to be (+1, -1, -1, -1) for the description of the
flat background Minkowski spacetime manifold so that $P_\mu Q^\mu = P_0 Q_0 - P_i Q_i$
where Greek indices $\mu, \nu, \eta .....= 0, 1, 2, 3$ and Latin indices
$i, j, k.....= 1, 2, 3$. The Levi-Cvita tensor is chosen to be $\varepsilon_{0123} = +1$
which obeys $\varepsilon^{\mu\nu\eta\kappa} \varepsilon_{\mu\nu\eta\kappa} = - 4!,
\varepsilon^{\mu\nu\eta\kappa} \varepsilon_{\mu\nu\eta\rho} = - 3! \delta^\kappa_\rho$, etc.
For the sake of brevity, we adopt the dot and cross products $R \cdot S = R^a S^a,
R \times S = f^{abc} R^a S^b T^c$ in the SU(N) Lie algebraic space where the generators
$T^a$ of the SU(N) Lie algebra  satisfy
the commutator $[T^a, T^b] = i f^{abc} T^c$ with $a, b, c ....= 1, 2, 3...........N^2 - 1$.}
where there is an explicit coupling between the 1-form non-Abelian gauge field,
an additional 1-form non-Abelian field  and the 2-form non-Abelian
auxiliary field [4], viz;
\begin{eqnarray}
{\cal L}_0 = - \frac{1}{4}\; F^{\mu\nu} \cdot F_{\mu\nu} + \frac{m^2}{2}\; \Phi^\mu \cdot \Phi_\mu
+ \frac{m}{4} \;\varepsilon^{\mu\nu\eta\kappa} \;{\cal F}_{\mu\nu} \cdot B_{\eta\kappa}.
\end{eqnarray}
Here the 2-form $F^{(2)} = d A^{(1)} + i A^{(1)} \wedge A^{(1)} \equiv
\frac{1}{2!} (dx^\mu \wedge dx^\nu) F_{\mu\nu} \cdot T$ defines  the curvature
tensor $F_{\mu\nu} = \partial_\mu A_\nu - \partial_\nu A_\mu - (A_\mu \times A_\nu)$
for the 1-form $(A^{(1)} = dx^\mu A_\mu \cdot T$) non-Abelian gauge field $A_\mu$ and
 ${\cal F}_{\mu\nu} = F_{\mu\nu} + f_{\mu\nu} - (A_\mu \times \Phi_\nu) - (\Phi_\mu \times A_\nu)$
 is the field strength tensor for the sum of potentials $A_\mu$ and $\Phi_\mu$. As a consequence,
 we note that $f_{\mu\nu} = \partial_\mu \Phi_\nu - \partial_\nu \Phi_\mu - (\Phi_\mu \times \Phi_\nu)$.
 It is self-evident that the 2-form $[B^{(2)} = \frac{1}{2!} (dx^\mu \wedge dx^\nu) B_{\mu\nu} \cdot T]$ field $B_{\mu\nu}$
 is an auxiliary field in the theory because it does not have a kinetic term.
 We take note of the fact that ($A_\mu, B_{\mu\nu}, \Phi_\mu$) have mass dimension one (in
 natural units) implying that $m$, present in the last term of (1), has the dimension of mass in 4D.

It is straightforward to check that the above Lagrangian density respects
($ \delta_g {\cal L}_0 = 0$) the following (1-form) YM gauge symmetry transformations ($\delta_g$)
\begin{eqnarray}
&& \delta_g A_\mu = D_\mu \Omega, \qquad \delta_g B_{\mu\nu} = - (B_{\mu\nu} \times \Omega), \qquad
\delta_g \Phi_\mu = - (\Phi_\mu \times \Omega), \nonumber\\
&& \delta_g F_{\mu\nu} = - (F_{\mu\nu} \times \Omega),\;\; \qquad \;\;\;\delta_g {\cal F}_{\mu\nu}
= - ({\cal F}_{\mu\nu} \times \Omega),
\end{eqnarray}
where $\Omega = \Omega \cdot T$ is the Lie-valued infinitesimal (Lorentz scalar) gauge parameter
and $D_\mu \Omega = \partial_\mu \Omega - (A_\mu \times \Omega)$ is the covariant derivative
w.r.t. the usual 1-form gauge field $A_\mu$. We note that the
antisymmetric tensor $f_{\mu\nu}$ does not transform covariantly under the (1-form) YM transformations $\delta_g$, namely;
\begin{eqnarray}
\delta_g f_{\mu\nu} = - (f_{\mu\nu} \times \Omega) + (\Phi_\mu \times \partial_\nu \Omega)
- (\Phi_\nu \times \partial_\mu \Omega),
\end{eqnarray}
which is essential for the covaraint transformation $\delta_g {\cal F}_{\mu\nu} = - ({\cal F}_{\mu\nu} \times \Omega)$.
We observe that $ \delta_g ({\cal F}_{\mu\nu} \cdot B_{\eta\kappa}) = 0$.
As a consequence,  all the terms in the Lagrangian density (1) are  {\it individually} gauge
invariant quantities.

There exists a local (2-form) tensor gauge symmetry in the theory. These transformations ($\delta_t$),
for all the relevant fields of the theory, are\footnote{Of course, we shall
{\it not} be exploiting these continuous symmetry transformations for our present discussions
on the derivation of  (anti-)BRST symmetry transformations within the framework of ``augmented''
Bonora-Tonin's superfield approach to BRST formalism.}
\begin{eqnarray}
\delta_t F_{\mu\nu} = \delta_t f_{\mu\nu} = \delta_t A_\mu = \delta_t \Phi_\mu = \delta_t {\cal F}_{\mu\nu} = 0,
 \;\qquad\;
\delta_t B_{\mu\nu} = \tilde D_\mu \Lambda_\nu - \tilde D_\nu \Lambda_\mu.
\end{eqnarray}
Here $\tilde D_\mu \Lambda_\nu = \partial_\mu \Lambda_\nu - (A_\mu \times \Lambda_\nu) - (\Phi_\mu \times \Lambda_\nu)$
is the covariant derivative w.r.t. the sum of 1-form fields $A_\mu$ and $\Phi_\mu$, and $\Lambda_\mu = \Lambda_\mu \cdot T$
is the infinitesimal Lorentz vector gauge transformation parameter. In fact, the
validity of the Bianchi identity $\tilde D_\mu {\cal F}_{\nu\eta}
+ \tilde D_\nu {\cal F}_{\eta\mu} + \tilde D_\eta {\cal F}_{\mu\nu} = 0$ is the root cause of the symmetry
invariance of the Lagrangian density (and corresponding action integral)
under the above local tensor gauge symmetry transformations. \\

\noindent
{\large \bf 3. (Anti-)BRST symmetries: superfield formalism}\\

\noindent
We discuss here separately the derivation of the nilpotent (anti-)BRST symmetry transformations
corresponding to the (1-form) YM gauge transformations (2) for (i) the non-Abelian
1-form gauge field $A_\mu$ and corresponding (anti-)ghost fields, and (ii) the
2-form non-Abelian field $B_{\mu\nu}$ and 1-form field $\Phi_\mu$. We also comment on the derivation of
CF condition and its importance.\\

\noindent
{\bf 3.1 (Anti-)BRST symmetries: 1-form gauge and ghost fields}\\

\noindent
We briefly recapitulate here the Bonora-Tonin's superfield approach to BRST formalism [12,13]
and derive the (anti-)BRST transformations for the 1-form field $A_\mu$ and (anti-)ghost fields
$(\bar C)C$ along with the Curci-Ferrari condition [11]. To this end in mind, first of all,
we generalize the curvature 2-form, defined in the language of Maurer-Cartan equation
$F^{(2)} = d A^{(1)} + i\; A^{(1)} \wedge A^{(1)}$ (and its innate operators),
onto the (4, 2)-dimensional supermanifold, as  [12]
\begin{eqnarray}
d & \to & \tilde d \;= \;dZ^M \partial_M \; \equiv \;dx^\mu \;\partial_\mu + d \theta \;\partial_\theta
+ d \bar \theta \;\partial_{\bar\theta}, \nonumber\\
A^{(1)} & \to & \tilde A^{(1)} = dZ^M \tilde A_M \equiv dx^\mu \tilde B_\mu (x,\theta,\bar\theta) + d \theta
\tilde {\bar F} (x,\theta,\bar\theta) + d \bar \theta \tilde F (x,\theta,\bar\theta), \nonumber\\
F^{(2)} &\to& \tilde F^{(2)} = \frac{1}{2!} (dZ^M \wedge dZ^N)\; \tilde F_{MN}
\equiv \tilde d \tilde A^{(1)} + i \;\tilde A^{(1)} \wedge \tilde A^{(1)},
\end{eqnarray}
where the superspace coordinates $Z^M = (x^\mu, \theta, \bar\theta)$ characterize the above (4, 2)-dimensional
supermanifold with spacetime variables $x^\mu$ (with $\mu = 0, 1, 2, 3$) and a pair of Grassmannian variables (with
$\theta^2 = \bar\theta^2 = 0, \theta \bar\theta + \bar\theta \theta = 0$). Corresponding superspace
derivatives are denoted by $\partial_M = (\partial_\mu, \partial_\theta, \partial_{\bar\theta})$.

The set of superfields ($\tilde B_\mu, \tilde F, \tilde {\bar F}$) form the super-multiplet
of a super vector field $\tilde A_M$ and these have the following super expansions [12]
\begin{eqnarray}
\tilde B_{\mu} (x, \theta, \bar\theta) &=& A_\mu (x) + \theta\; \bar R_\mu (x) + \bar \theta\; R_\mu (x)
+ i \;\theta \;\bar\theta \; S_\mu (x), \nonumber\\
\tilde F (x, \theta, \bar\theta) &=& C (x) + i\;\theta\; \bar B_1 (x) + i\;\bar \theta\; B_1 (x)
+ i \;\theta\; \bar\theta \; s (x), \nonumber\\
\tilde {\bar F} (x, \theta, \bar\theta) &=& \bar C (x) +
 i\;\theta\; \bar B_2 (x) + i\;\bar \theta\; B_2 (x)
+ i \;\theta \;\bar\theta \; \bar s (x),
\end{eqnarray}
in terms of the basic fields ($A_\mu, C, \bar C$) of the BRST invariant non-Abelian 1-form
theory [21] and secondary fields $R_\mu, \bar R_\mu, S_\mu, B_1, \bar B_1, B_2, \bar B_2, s, \bar s$.
The secondary fields are determined in terms of the basic and auxiliary fields of the 4D non-Abelian
1-form theory by imposing the celebrated horizontality condition (HC) which states that all the
Grassmannian components of the super 2-form  $\tilde F^{(2)} = \frac{1}{2!} (dZ^M \wedge dZ^N) \tilde F_{MN}$
should be set equal to zero (see, e.g., [12]). The explicit computation of $\tilde F^{(2)}$ and
application of HC, lead to
\begin{eqnarray}
&& R_\mu = D_\mu C,\; \quad \bar R_\mu = D_\mu \bar C,\; \quad B_1 = - \frac{i}{2} (C \times C),\;\quad
\bar B_2 = -  \frac{i}{2} (\bar C \times \bar C), \nonumber\\
&& S_\mu = D_\mu B \;+ \;i \;(D_\mu C \times \bar C) \equiv - D_\mu \bar B \;- \;i \;(D_\mu \bar C \times C), \nonumber\\
&& s = - (\bar B \times C),\; \qquad \bar s = + (B \times \bar C),\; \qquad B + \bar B = -i (C \times \bar C),
\end{eqnarray}
where we have identified $B_2 = B, \bar B_1 = \bar B$ to be consistent with the  Nakanishi-Lautrup notations
of the auxiliary fields in the 4D non-Abelian 1-form theory. We also note here that we have already derived
the celebrated Curci-Ferrari (CF) condition (i.e. $B + \bar B = -i (C \times \bar C)$) for the non-Abelian theory [21].

With the above substitutions, we obtain the following super expansions
of the above multiplet superfields\footnote{The superscript ${(h)}$ on the multiplet superfields denotes
the fact that the expansions have been obtained after the application of celebrated HC.}
 (after the application of the HC) [12]
\begin{eqnarray}
\tilde B^{(h)}_{\mu} (x, \theta, \bar\theta) &=& A_\mu (x) + \theta\; \bar D_\mu \bar C (x) + \bar \theta\; D_\mu C (x) \nonumber\\
&+&  \theta \;\bar\theta \; [i D_\mu  C - (D_\mu C \times \bar C)] (x) \nonumber\\
&\equiv& A_\mu (x) + \theta\; (s_{ab} A_\mu (x)) + \bar \theta\; (s_b A_\mu (x)) + \theta\; \bar\theta\;
(s_b s_{ab} A_\mu (x)), \nonumber\\
\tilde F^{(h)} (x, \theta, \bar\theta) &=& C (x) + \theta\; (i \bar B (x)) + \bar \theta\; \Bigl [ \frac{1}{2} (C \times C) (x) \Bigr ]
+  \theta \bar\theta \; [-i (\bar B \times C)(x)] \nonumber\\
&\equiv& C (x) + \theta\; (s_{ab} C (x)) + \bar \theta\; (s_b C (x)) + \theta\; \bar\theta\;
(s_b s_{ab} C (x)), \nonumber\\
\tilde {\bar F}^{(h)} (x, \theta, \bar\theta) &=& \bar C (x) +
\theta\; \Bigl [\frac{1}{2} ( \bar C \times \bar C) (x) \Bigr ] + \bar \theta\; (i \bar B (x))
+ \theta \bar\theta \; [(+ i (B \times \bar C) (x)] \nonumber\\
&\equiv& \bar C (x) + \theta\; (s_{ab} \bar C (x)) + \bar \theta\; (s_b \bar C (x)) + \theta\; \bar\theta\;
(s_b s_{ab} \bar C (x)),
\end{eqnarray}
which define the (anti-)BRST symmetry transformations $s_{(a)b}$ for the 1-form non-Abelian gauge field and
corresponding (anti-)ghost fields as
\begin{eqnarray}
&& s_b A_\mu = D_\mu C,\; \quad s_b C = \frac{1}{2} (C \times C),\; \quad
s_b \bar B = - (\bar B \times C),\; \quad s_b B = 0, \nonumber\\
&& s_{ab} A_\mu = D_\mu \bar C, \quad s_{ab} \bar C = \frac{1}{2} (\bar C \times \bar C), \; \;
s_{ab} B = - (B \times C),\; \; s_{ab} \bar B = 0.
\end{eqnarray}
We note, from equation (8), that $s_b \leftrightarrow \mbox{Lim}_{\theta \to 0} (\partial/\partial \bar\theta),
s_{ab} \leftrightarrow \mbox{Lim}_{\bar\theta \to 0} (\partial/\partial \theta)$.
It is worthwhile to mention that the requirements of the off-shell nilpotency ($s_{(a)b}^2 = 0$)
and absolute anticommutativity ($s_b s_{ab} + s_{ab} s_b = 0$) of the above (anti-)BRST symmetry
transformations leads to the derivation of the nilpotent (anti-)BRST  transformations for the
Nakanishi-Lautrup
auxiliary fields as: $s_b B = 0, s_b \bar B = - (\bar B \times C), s_{ab} \bar B = 0, s_{ab} B =
- (B \times \bar C)$, etc.

We wrap up this subsection with the following remarks. First, the absolute anticommutatvity
$\{s_b, s_{ab} \} A_\mu = 0$ is true if and only if the CF condition ($B + \bar B + i\; (C \times \bar C) = 0$)
is satisfied. Second, it can
be checked that CF condition is (anti-)BRST invariant (i.e. $s_{(a)b} [B + \bar B + i\; (C \times \bar C)]
= 0$) quantity and, therefore, this condition is physical (in some sense). Finally, spacetime component
of the super 2-form $\tilde F^{(2)}$ leads to the derivation of the following
super expansion of the antisymmetric super curvature [12]:
\begin{eqnarray}
\tilde F^{(h)}_{\mu\nu} (x,\theta,\bar\theta) &=& F_{\mu\nu} - \theta \;(F_{\mu\nu} \times \bar C)
- \;\bar\theta \;(F_{\mu\nu} \times C) \nonumber\\
&+& \theta\; \bar\theta\; \bigl [(F_{\mu\nu} \times C) \times \bar C - i\; F_{\mu\nu} \times B \bigr ],
\end{eqnarray}
which implies the following
nilpotent and anticommuting (anti-)BRST symmetry transformations for it (i.e. the curvature tensor), namely;
\begin{eqnarray}
&& s_b F_{\mu\nu} = - (F_{\mu\nu} \times C),\; \qquad \;s_{ab} F_{\mu\nu} = - (F_{\mu\nu} \times \bar C), \nonumber\\
&& s_b s_{ab} F_{\mu\nu} = (F_{\mu\nu} \times C) \times \bar C - \;i\; F_{\mu\nu} \times B.
\end{eqnarray}
We observe that $- \frac{1}{4} \tilde F^{\mu\nu (h)} (x,\theta,\bar\theta) \cdot
\tilde F^{(h)}_{\mu\nu} (x,\theta,\bar\theta) = - \frac{1}{4} F^{\mu\nu} (x)  \cdot F_{\mu\nu} (x)$.
The Grassmannian independence of the l.h.s. implies that
the kinetic  term remains invariant under the above (anti-)BRST symmetry transformations
in view of the fact that $s_b \leftrightarrow \mbox{Lim}_{\theta \to 0} (\partial/\partial \bar\theta),
s_{ab} \leftrightarrow \mbox{Lim}_{\bar\theta \to 0} (\partial/\partial \theta)$. \\

\noindent
{\bf 3.2 (Anti-)BRST symmetries: 1-form field $\Phi_\mu$ and 2-form field $B_{\mu\nu}$}\\

\noindent
As pointed out earlier, it can be checked that there are useful combinations of fields
that are gauge-invariant quantities under the 1-form YM transformations (2). To be
specific, it can be verified that
\begin{eqnarray}
\delta_g (F_{\mu\nu} \cdot B_{\eta\kappa}) = 0,\;\;\; \qquad\;\;\;  \delta_g (F_{\mu\nu} \cdot \Phi_{\eta}) = 0.
\end{eqnarray}
As a consequence, these quantities are physical objects as far as the gauge transformations (2)
are concerned. Thus, we demand that these quantities should remain independent of the Grassmannian
variables when they are generalized onto the (4, 2)-dimensional supermanifold. In other words,
we invoke the following gauge invariant restrictions (GIRs) on the (super)fields:
\begin{eqnarray}
&&\tilde F_{\mu\nu}^{(h)} (x,\theta,\bar\theta) \cdot \tilde B_{\eta\kappa} (x,\theta,\bar\theta)
= F_{\mu\nu} (x) \cdot B_{\eta\kappa} (x), \nonumber\\
&&\tilde F_{\mu\nu}^{(h)} (x,\theta,\bar\theta) \cdot \tilde \Phi_\eta (x,\theta,\bar\theta)
= F_{\mu\nu} (x) \cdot \Phi_{\eta} (x).
\end{eqnarray}
Physically, the above requirements imply the (anti-)BRST invariance of the gauge
invariant quantities listed in (12). The above choice is important because, with the
inputs from the super expansion in (10), we shall be able to obtain the (anti-)BRST
transformations for the fields $\Phi_\mu$ and $B_{\mu\nu}$.

Towards above goal in mind,
let us exploit the following general expansions for the superfields in the
GIRs, listed in (13), namely;
\begin{eqnarray}
\tilde B_{\mu\nu} (x,\theta,\bar\theta) &=& B_{\mu\nu} (x) + \theta \; \bar R_{\mu\nu} (x)
+ \bar\theta\; R_{\mu\nu} (x) + i \;\theta \;\bar\theta \;S_{\mu\nu} (x), \nonumber\\
\tilde \Phi_\mu (x,\theta,\bar\theta) &=& \Phi_\mu (x) + \theta\; \bar S_\mu (x) + \bar\theta\;
S_\mu (x) + i \;\theta \;\bar\theta\; T_\mu (x),
\end{eqnarray}
where $(\Phi_\mu, T_\mu, B_{\mu\nu},S_{\mu\nu}$) are the bosonic (even) fields and
$(S_\mu, \bar S_\mu, R_{\mu\nu}, \bar R_{\mu\nu})$ are the fermionic (odd) fields in
the above super expansion. The secondary fields $S_\mu, \bar S_\mu, T_\mu, R_{\mu\nu},
\bar R_{\mu\nu}, S_{\mu\nu}$ are to be determined from the GIRs given in (13). In fact,
with the help of (13) and (14), we have the following
\begin{eqnarray}
&& R_{\mu\nu} = - (B_{\mu\nu} \times C),\; \qquad\; \bar R_{\mu\nu} = - (B_{\mu\nu} \times \bar C), \nonumber\\
&& S_{\mu\nu} = - i \bigl [ (B_{\mu\nu} \times C) \times \bar C - i \; (B_{\mu\nu} \times B) \bigr ], \nonumber\\
&& S_\mu = - (\Phi_\mu \times C), \;\qquad \;\bar S_\mu = - (\Phi_\mu \times \bar C), \nonumber\\
&& T_\mu = - i \bigl [(\Phi_\mu \times C) \times \bar C - i \;(\Phi_\mu \times B) \bigr ].
\end{eqnarray}
The substitution of the above expressions in (14) leads to\footnote{It can be checked that
$\tilde \Phi^{\mu(g)} (x,\theta,\bar\theta) \cdot \tilde \Phi^{(g)}_\mu (x,\theta,\bar\theta) = \Phi_\mu (x) \cdot \Phi^\mu (x)$ and
$\tilde B^{\mu\nu(g)} (x,\theta,\bar\theta) \cdot \tilde B^{(g)}_{\mu\nu} (x,\theta,\bar\theta) = B^{\mu\nu} (x) \cdot B_{\mu\nu} (x)$
which demonstrate the gauge and (anti-)BRST invariance of ($\Phi^\mu \cdot \Phi_\mu$) and ($B^{\mu\nu} \cdot B_{\mu\nu}$) in
view of $s_b \leftrightarrow \mbox{Lim}_{\theta \to 0} (\partial/\partial \bar\theta),
s_{ab} \leftrightarrow \mbox{Lim}_{\bar\theta \to 0} (\partial/\partial \theta)$.}
\begin{eqnarray}
\tilde B^{(g)}_{\mu\nu} (x,\theta,\bar\theta) &=& B_{\mu\nu} - \theta \; ( B_{\mu\nu} \times C) (x)
+ \bar\theta\; (B_{\mu\nu} \times \bar C) (x) \nonumber\\
&+& \theta \;\bar\theta \;
[(B_{\mu\nu} \times C) \times \bar C - i \;(B_{\mu\nu} \times B)] (x), \nonumber\\
\tilde \Phi^{(g)}_\mu (x,\theta,\bar\theta) &=& \Phi_\mu (x) - \theta\; (\Phi_\mu \times \bar C) (x) + \bar\theta\;
(\Phi_\mu \times C) (x) \nonumber\\
& +& \theta \;\bar\theta\; [(\Phi_\mu \times C) \times \bar C - i\; (\Phi_\mu \times B) ](x),
\end{eqnarray}
where the superscript $(g)$ on the superfields denotes that the superfields have been obtained
after the application of the GIRs (listed in (13)).
Taking the inputs from our previous discussions, it is clear that the off-shell nilpotent
($s_{(a)b}^2 = 0$) (anti-)BRST transformations for the fields $\Phi_\mu$ and $B_{\mu\nu}$ are
\begin{eqnarray}
&& s_b \Phi_\mu = - (\Phi_\mu \times C),\; \quad s_b B_{\mu\nu} = - (B_{\mu\nu} \times C), \nonumber\\
&& s_b s_{ab} \Phi_\mu = (\Phi_\mu \times C) \times \bar C - i \;(\Phi_\mu \times B), \nonumber\\
&& s_{ab} \Phi_\mu = - (\Phi_\mu \times \bar C),\; \quad s_{ab} B_{\mu\nu} = -  B_{\mu\nu} \times \bar C), \nonumber\\
&& s_b s_{ab} B_{\mu\nu} = (B_{\mu\nu} \times C) \times \bar C - i \; B_{\mu\nu} \times B.
\end{eqnarray}
Thus, the set of transformations (9) and (17) produce all the off-shell nilpotent (anti-)BRST
symmetry transformations for {\it all} the basic fields of our present 4D topologically massive gauge theory.

We have seen earlier that the curvature tensor ${\cal F}_{\mu\nu}$ transforms covariantly under the
(1-form) YM gauge transformations (2) (i.e. $\delta_g {\cal F}_{\mu\nu} = - ({\cal F}_{\mu\nu} \times \Omega)$).
Its transformations under the (anti-)BRST symmetry transformations can also be obtained within the framework of
``augmented'' BT superfield formalism. This can be obtained by plugging in the explicit expansions of $\tilde B_\mu^{(h)},
\tilde \Phi_\mu^{(g)}, \tilde F_{\mu\nu}^{(h)}, \tilde f_{\mu\nu}^{(g)}$ in the following expression for
this super curvature tensor
\begin{eqnarray}
\tilde {\cal F}^{(g, h)}_{\mu\nu} (x,\theta,\bar\theta) = \tilde F_{\mu\nu}^{(h)} + \tilde f_{\mu\nu}^{(g)}
- (\tilde B_\mu^{(h)} \times \tilde \Phi_\nu^{(g)}) + (\tilde B_\nu^{(h)} \times \tilde \Phi_\mu^{(g)}),
\end{eqnarray}
where the explicit expression for $\tilde f_{\mu\nu}^{(g)}$ is as given below
\begin{eqnarray}
&& \tilde f_{\mu\nu}^{(g)} (x,\theta,\bar\theta) = \partial_\mu \tilde \Phi_\nu^{(g)} 
- \partial_\nu \tilde \Phi_\mu^{(g)}
- (\tilde \Phi^{(g)}_\mu \times \tilde \Phi_\nu^{(g)}) \nonumber\\
&&\equiv f_{\mu\nu} (x) + \theta\; (s_{ab} f_{\mu\nu} (x)) + \bar\theta\; (s_b f_{\mu\nu} (x))
+ \theta\; \bar\theta\; (s_b s_{ab} f_{\mu\nu} (x)).
\end{eqnarray}
In the above, the explicit forms of the nilpotent (anti-)BRST symmetry transformations for the curvature tensor $f_{\mu\nu}$ are
listed below
\begin{eqnarray}
&&s_b f_{\mu\nu} = - (f_{\mu\nu} \times C) + (\Phi_\mu \times \partial_\nu C)
- (\Phi_\nu \times \partial_\mu C), \nonumber\\
&& s_{ab} f_{\mu\nu} = - (f_{\mu\nu} \times \bar C)
+ (\Phi_\mu \times \partial_\nu \bar C)
- (\Phi_\nu \times \partial_\mu \bar C), \nonumber\\
&& s_b s_{ab} f_{\mu\nu} = (f_{\mu\nu} \times C) \times \bar C - i \;(f_{\mu\nu} \times B) \nonumber\\
&&- (\Phi_\mu \times \partial_\nu C) \times \bar C + (\Phi_\nu \times C) \times \partial_\mu \bar C \nonumber\\
&& + (\Phi_\nu \times \partial_\mu C) \times \bar C - (\Phi_\mu \times C) \times \partial_\nu \bar C
+ i\; (\Phi_\mu - \Phi_\nu) \times B.
\end{eqnarray}
The substitutions of the expressions in (10), (16), (19), (20), with a little dose of algebra, leads to the following
expansion for the super curvature tensor\footnote{It is interesting to check that $\tilde {\cal F}^{(g, h)}_{\mu\nu} (x,\theta,\bar\theta)
\cdot \tilde B^{(g)}_{\eta\kappa} (x,\theta,\bar\theta) = {\cal F}_{\mu\nu} (x) \cdot B_{\eta\kappa} (x)$. This shows the
(anti-)BRST invariance of the topological massive term of the Lagrangian density (1).}
\begin{eqnarray}
\tilde {\cal F}^{(g, h)}_{\mu\nu} (x, \theta, \bar\theta) &=&  {\cal F}_{\mu\nu} (x) - \theta\; ({\cal F}_{\mu\nu} \times \bar C) (x)
- \bar\theta\; ({\cal F}_{\mu\nu} \times C) (x) \nonumber\\
&+& \theta\;\bar\theta\; \bigl [({\cal F}_{\mu\nu} \times C) \times \bar C - i\; ({\cal F}_{\mu\nu} \times B) \bigr ] (x).
\end{eqnarray}
The above expansion leads to the derivation of the (anti-)BRST symmetry transformations for the curvature tensor
${\cal F}_{\mu\nu}$ as given below
\begin{eqnarray}
&&s_b {\cal F}_{\mu\nu} = - ({\cal F}_{\mu\nu} \times C),\; \qquad s_{ab} {\cal F}_{\mu\nu}
= - ({\cal F}_{\mu\nu} \times \bar C), \nonumber\\
&& s_b s_{ab} {\cal F}_{\mu\nu} = ({\cal F}_{\mu\nu} \times C) \times \bar C - \;i \;({\cal F}_{\mu\nu} \times B).
\end{eqnarray}
Thus, we have derived all the proper (anti-)BRST symmetry transformations for all the dynamical
fields and their curvature tensors (that are present in the theory) by exploiting the
``augmented'' BT superfield approach to BRST formalism where the HC and GIRs blend together
in a meaningful manner.\\

\noindent
{\bf 3.3 Curci-Ferrari condition: absolute anticommutatvity of the nilpotent (anti-)BRST
symmetries  and coupled Lagrangian densities}\\

\noindent
One of the key signatures of a $p$-form ($p = 1, 2, 3.....$)
gauge theory is the existence of the first-class constraints
in the language of Dirac's prescription for the classification scheme [22,23].
When these theories are discussed within the framework of BRST formalism,
there always exists (one or more number of) Curci-Ferrari (CF) type conditions. For instance,
in the simplest case of an Abelian 1-form (anti-)BRST invariant gauge theory, we
have a trivial CF type condition $B + \bar B = 0$ as can be seen from
the Abelian limit of the CF condition (i.e. $B + \bar B + i \;(C \times \bar C) = 0$)
mentioned in (7).

The beauty of the Bonora-Tonin's (BT) superfield approach to BRST formalism is
that the above CF-type restrictions emerge very naturally\footnote{In fact, it
turns out that when the coefficient of the differential ($d\theta \wedge d\bar\theta$)
of the super curvature 2-form $\tilde F^{(2)} = \frac{1}{2!} (dZ^M \wedge dZ^N) \tilde F_{MN}$ is
set equal to zero due to HC, we obtain the CF condition $B + \bar B + i (C \times \bar C) = 0$ within
the framework of BT superfield formalism.}.
They are always (anti-)BRST invariant as has been proven within the framework of ``augmented''
BT superfield approach to the non-Abelian TMGT where the
dynamical 2-form gauge field is coupled with the 1-form gauge field through
the celebrated $B \wedge F$ term (see, e.g. [18] for details).
In fact, the existence of CF type restrictions are responsible for the
absolute anticommutativity of the (anti-)BRST symmetry transformations in the context
of any arbitrary $p$-form gauge theory, discussed within the framework of BRST formalism. For instance,
it can be explicitly checked that, in our present topologically massive theory, the following
anticommutators
\begin{eqnarray}
\{ s_b, s_{ab} \}\; A_\mu = 0, \;\qquad\;\;
\{ s_b, s_{ab} \}\; \Phi_\mu = 0,\;   \qquad \;\; \{ s_b, s_{ab} \} \;B_{\mu\nu} = 0,
\end{eqnarray}
are true if and only if we take into account the (anti-)BRST invariant
CF condition $B + \bar B + i \;(C \times \bar C) = 0$.
This anticommutativity property is valid for all the {\it rest} of the fields
(of our present 4D topologically massive theory) {\it without}
invoking the above CF type restriction in any form.

Another important contribution of the CF type restrictions is the derivation of the
coupled (but equivalent) Lagrangian densities for a given $p$-form gauge theory. In the case of a simple
Abelian 1-form gauge theory, these coupled Lagrangian densities merge into a single Lagrangian density. For our
case, it can be checked that the following Lagrangian densities\footnote{It
should be noted that, for the 4D (anti-)BRST invariant theory, the terms inside the
square brackets are unique in the sense that these are the only combinations that have mass
dimensions two (in natural units) and ghost number equal to zero.}
\begin{eqnarray}
{\cal L}_B &=& {\cal L}_0 + s_b s_{ab} \Bigl [\frac{i}{2} A_\mu \cdot A^\mu
+ \frac{1}{4} B_{\mu\nu} \cdot B^{\mu\nu} + \frac{i}{2} \Phi_\mu \cdot \Phi^\mu + \bar C \cdot C \Bigr ], \nonumber\\
{\cal L}_{\bar B} &=& {\cal L}_0 - s_{ab} s_b \Bigl [\frac{i}{2} A_\mu \cdot A^\mu
+ \frac{1}{4} B_{\mu\nu} \cdot B^{\mu\nu} + \frac{i}{2} \Phi_\mu \cdot \Phi^\mu + \bar C \cdot C \Bigr ],
\end{eqnarray}
are found to respect the (anti-)BRST symmetry transformations on a surface in the 4D spacetime
manifold where the CF condition $B + \bar B + i\; (C \times \bar C) = 0$ is satisfied. In fact,
taking the help of transformations (9) and (17), we can derive the above Lagrangian densities
explicitly as given below (see, e.g. [21])
\begin{eqnarray}
{\cal L}_B &=& {\cal L}_0 + B \cdot (\partial_\mu A^\mu) + \frac{1}{2} (B \cdot B + \bar B \cdot \bar B)
- i\; \partial_\mu \bar C \cdot D^\mu C, \nonumber\\
{\cal L}_{\bar B} &=& {\cal L}_0 - \bar B \cdot (\partial_\mu A^\mu) + \frac{1}{2} (B \cdot B + \bar B \cdot \bar B)
- i \;D_\mu \bar C \cdot \partial^\mu C.
\end{eqnarray}
It is worthwhile to mention that the (1-form) YM gauge (and (anti-)BRST) invariance of ($B^{\mu\nu} \cdot B_{\mu\nu}$)
and ($\Phi^\mu \cdot \Phi_\mu$) in (24) imply that there are no gauge-fixing and Faddeev-Popov ghost terms
for the fields $B_{\mu\nu}$ and $\Phi_\mu$ that could be incorporated in (25).
The equivalence of the above Lagrangian densities can be easily checked by the following equality
\begin{eqnarray}
B \cdot (\partial_\mu A^\mu) - i \partial_\mu \bar C \cdot D^\mu C =
- \bar B \cdot (\partial_\mu A^\mu) - i D_\mu \bar C \cdot \partial^\mu C,
\end{eqnarray}
which is valid only on a surface in the 4D Minkowski spacetime manifold that is
described by the field equation $B + \bar B + i\; (C \times \bar C) = 0$. In other
words, the above (anti-)BRST invariant CF condition is responsible for the derivation of the (anti-)BRST
invariant coupled
Lagrangian densities (25).

As a closing remark to this subsection, we would like to point out that there is yet another
way to derive the CF condition besides the BT method of superfield formalism which produces it
very naturally [12]. For instance, one can derive the following Euler-Lagrange equations of motion
for the 1-form gauge field from the coupled Lagrangian densities given in (25), namely;
\begin{eqnarray}
D_\mu F^{\mu\nu} + \frac{m}{2} \;\varepsilon^{\nu\mu\eta\kappa} \;\tilde D_\mu B_{\eta\kappa}
- \partial^\nu B &=& + \;i \;(\partial^\nu \bar C \times C), \nonumber\\
D_\mu F^{\mu\nu} + \frac{m}{2} \;\varepsilon^{\nu\mu\eta\kappa}\; \tilde D_\mu B_{\eta\kappa}
+ \partial^\nu \bar B &=& -\; i \;(\bar C \times \partial^\nu C).
\end{eqnarray}
If we take the difference of the above equations, we easily obtain the CF condition $B + \bar B +
i \;(C \times \bar C) = 0$. We conclude, ultimately, that the (anti-)BRST invariant
CF condition is hidden in the coupled
(but equivalent) Lagrangian densities (25) of our present theory, in a subtle manner. \\

\noindent
{\large \bf 4. BRST charge as the generator: a new feature}\\

\noindent
Let us focus on the BRST invariant Lagrangian density ${\cal L}_B$ (cf. (25)). It can
be checked that this Lagrangian density transforms to a total spacetime derivative
(i.e. $s_b {\cal L}_B = \partial_\mu [B \cdot D^\mu C]$) under the BRST
transformations $s_b$ (cf. (9),(11),(17),(22)). This symmetry invariance can be captured
within the framework of ``augmented'' BT superfield formalism because, the super Lagrangian density
of our present theory (i.e.
an analogue of ${\cal L}_B$) is given by
\begin{eqnarray}
&&\tilde {\cal L}_B = - \frac{1}{4} \tilde F^{\mu\nu(h)} \cdot \tilde F^{(h)}_{\mu\nu}
+ \frac{m^2}{2} \tilde \Phi^{\mu(g)} \cdot \tilde \Phi^{(g)}_\mu
+ \frac{m}{4} \;\varepsilon^{\mu\nu\eta\kappa}\; 
\tilde {\cal F}^{(g,h)}_{\mu\nu} \cdot \tilde B^{(g)}_{\eta\kappa}
\nonumber\\
&&+ \frac{\partial}{\partial\bar\theta}\;\frac{\partial}{\partial \theta} \Bigl [ \;\frac{i}{2}\;
\tilde B^{\mu(h)} \cdot \tilde B^{(h)}_\mu + \frac{1}{4} \;\tilde B^{\mu\nu(g)} \cdot \tilde B^{(g)}_{\mu\nu}
\nonumber\\
&& + \frac{i}{2} \;\tilde \Phi^{\mu(g)} \cdot \tilde \Phi^{(g)}_\mu \;+ \;\tilde {\bar F}^{(h)} \cdot \tilde F^{(h)} \Bigr ],
\end{eqnarray}
where the expressions for superfields have been taken after the applications of HC and GIRs. It is now straightforward
to note that the following mapping exists between the ordinary 4D symmetry and superfield formalism:
\begin{eqnarray}
\frac{\partial}{\partial \bar\theta} \bigl [\tilde {\cal L}_B\bigr ] = 0 \qquad \Leftrightarrow \qquad
s_b {\cal L}_B = \partial_\mu [B \cdot D^\mu C].
\end{eqnarray}
The above mapping is true due to the fact that (i) all the first three terms of the super
Lagrangian density are effectively independent of the Grassmannian variables $\theta$ and $\bar\theta$,
(ii) there exists a  relationship $s_b \leftrightarrow \mbox{Lim}_{\theta \to 0} (\partial/\partial\bar\theta)$, and
(iii) the translation operator $(\partial/\partial\bar\theta)$
along the $\bar\theta$-direction of the (4, 2)-dimensional supermanifold
is nilpotent of order two (i.e. $(\partial/\partial\theta)^2 = 0$).

As a consequence of the above continuous symmetry invariance, we can use the following Noether formula for the current
in terms of the generic field $\Psi_i = A_\mu, \Phi_\mu, B_{\mu\nu}, C, \bar C, B, \bar B$, viz;
\begin{eqnarray}
J^\mu_{(b)} = (s_b \Psi_i) \; \cdot \Bigl (\frac{\partial {\cal L}_B} {\partial_\mu \Psi_i}\Bigr) - B \cdot D^\mu C,
\end{eqnarray}
which leads to the derivation of conserved Noether current $J^\mu_{(b)}$ as
\begin{eqnarray}
J^\mu_{(b)} &=& B \cdot D^\mu C - \bigl [F^{\mu\nu} - \frac{m}{2}\; \varepsilon^{\mu\nu\eta\kappa}\;
 B_{\eta\kappa} \bigr ] \cdot D_\nu C \nonumber\\
&-& \frac{m}{2}\; \varepsilon^{\mu\nu\eta\kappa}\; (\Phi_\nu \times C) \cdot B_{\eta\kappa}
+ \frac{i}{2}\; \partial^\mu \bar C \cdot (C \times C).
\end{eqnarray}
The above expression can be recast in the following (more readable) form
\begin{eqnarray}
J^\mu_{(b)} &=& B \cdot D^\mu C - \partial^\mu B \cdot C - \frac{i}{2} (\partial^\mu \bar C \times C) \cdot C\nonumber\\
&+& \partial_\nu \Bigl [(F^{\nu\mu}
+ \frac{m}{2} \varepsilon^{\mu\nu\eta\kappa} B_{\eta\kappa}) \cdot C \Bigr ],
\end{eqnarray}
by exploiting the following set of Euler-Lagrange equations of motion
\begin{eqnarray}
&& D_\mu \;F^{\mu\nu} + \frac{m}{2}\; \varepsilon^{\nu\mu\eta\kappa}\; \tilde D_\mu\; B_{\eta\kappa}
- \partial^\nu B = + \;i \;(\partial^\nu \bar C \times C),  \nonumber\\
&& \partial_\mu (D^\mu C) = 0,\; \qquad D_\mu (\partial^\mu \bar C) = 0,\; \qquad 
F^{\mu\nu} = m \;\varepsilon^{\mu\nu\eta\kappa}\; B_{\eta\kappa},
\nonumber\\
&& \varepsilon^{\mu\nu\eta\kappa}\; \tilde D_\nu\; B_{\eta\kappa} + 2\; m \;\Phi^\mu = 0,
\; \qquad  \;\;\;{\cal F}_{\mu\nu} = 0,
\end{eqnarray}
which emerge from the Lagrangian density ${\cal L}_B$. It should be recalled that we have
taken $\tilde D_\mu B_{\nu\eta} = \partial_\mu B_{\eta\kappa} - (A_\mu \times B_{\eta\kappa}) - (\Phi_\mu \times B_{\eta\kappa})$.
The conservation law $\partial_\mu J^\mu_{(b)} = 0$ can also be proven by exploiting the above equations.

The conserved current $J^\mu_{(b)}$ leads to the derivation of conserved charge $Q_b = \int d^3 x J^0_{(b)}$.
The explicit expression for this charge is
\begin{eqnarray}
Q_b = \int d^3 x\; \bigl [B \cdot D^0 C - \dot B \cdot C - \frac{i}{2}\; \dot {\bar C} \cdot (C \times C) \bigr ],
\end{eqnarray}
where we have dropped the total space derivative terms because of the Gauss divergence theorem. The above expression for the
BRST charge is exactly same in appearance as the charge in the case of self-interacting non-Abelian 1-form gauge theory.
However, there is a key difference because, in the above, $\dot B$ has the explicit form (due to Euler-lagrange equation
of motion) as
\begin{eqnarray}
\dot B \equiv \partial^0 B = D_i F^{i0} + \frac{m}{2}\; \varepsilon^{0ijk} \tilde D_i B_{jk} - i\; (\dot {\bar C} \times C),
\end{eqnarray}
which contains the basic fields ($A_\mu, \Phi_\mu, C, \bar C$) as well as auxiliary field $B_{\mu\nu}$ (and their derivatives).
Only in the limits $B_{\mu\nu} \to 0, \Phi_\mu \to 0$ does the expression for the above charge reduces to the case of
non-Abelian 1-form theory.

It is worthwhile to point out that the expression for $Q_b$ contains the canonical momenta for all the dynamical
fields of our present theory. However, as is evident, there is no momentum for the $B_{\mu\nu}$ field in the expression for
the above charge. Thus, the following general formula for the BRST symmetry transformations (in terms of $Q_b$), namely;
\begin{eqnarray}
s_b \;\Phi_i = -\; i\; \Bigl [\Psi_i, Q_b \Bigr ]_{(\pm)}, \qquad \Psi_i = A_0, A_i, C, \bar C, \Phi_0, \Phi_i,
\end{eqnarray}
where $(\pm)$ signs on the square bracket correspond to the (anti)commutators for the generic field $\Psi_i$ being
(fermionic)bosonic in nature, is capable of producing the BRST transformations for all the {\it dynamical} fields. However,
it is interesting to point out that it fails, for the obvious reasons, to generate the BRST transformation
(i.e. $s_b B_{\mu\nu} = - (B_{\mu\nu} \times C)$) for the field $B_{\mu\nu}$.

It is worthwhile to state that, besides $B_{\mu\nu}$ field, there are other auxiliary fields in the theory. These
are nothing but the Nakanishi-Lautrup (NL) type auxiliary fields. However, there is a distinct difference between the above
cited auxiliary
fields. As pointed out after equation (9), the requirements of the nilpotency and
anticommutativity of the (anti-)BRST symmetry transformation $s_{(a)b}$ produce the (anti-)BRST symmetry transformations
for the NL type auxiliary fields $B$ and $\bar B$. However, even these sacrosanct requirements of the BRST formalism do {\it not}
produce the BRST transformation for the $B_{\mu\nu}$ field.
This is a novel feature of our present theory which has not been observed in the application of the BRST formalism
to 4D (non-)Abelian 1-form and Abelian 2-form and 3-form gauge theories [21,16,17].
Thus, ultimately, we note that there is a key difference between the
NL type auxiliary fields and the auxiliary field $B_{\mu\nu}$ of our present theory.\\

\noindent
{\large \bf 5. Anti-BRST charge: as the symmetry generator}\\

\noindent
It can be checked from the anti-BRST symmetry transformations (listed in equations (9), (11), (17) and (22))
that the Lagrangian density ${\cal L}_{\bar B}$ of (25), transforms to a total spacetime derivative
(i.e. $s_{ab} {\cal L}_{\bar B} = - \partial_\mu [ \bar B \cdot D^\mu \bar C]$) under these transformations.
This anti-BRST invariance can be written in terms of the superfields, obtained after the application of HC and GIRs,
within the framework of the ``augmented'' BT superfield formalism. Towards this goal in mind, let us express the Lagrangian
density ${\cal L}_{\bar B}$ as
\begin{eqnarray}
&&\tilde {\cal L}_B = - \frac{1}{4} \tilde F^{\mu\nu(h)} \cdot \tilde F^{(h)}_{\mu\nu}
+ \frac{m^2}{2} \tilde \Phi^{\mu(g)} \cdot \tilde \Phi^{(g)}_\mu
+ \frac{m}{4} \;\varepsilon^{\mu\nu\eta\kappa} \;\tilde {\cal F}^{(g,h)}_{\mu\nu} \cdot \tilde B^{(g)}_{\eta\kappa}
\nonumber\\
&&- \frac{\partial}{\partial\theta}\;\frac{\partial}{\partial \bar \theta} \Bigl [ \;\frac{i}{2}\;
\tilde B^{\mu(h)} \cdot \tilde B^{(h)}_\mu + \frac{1}{4} \;\tilde B^{\mu\nu(g)} \cdot \tilde B^{(g)}_{\mu\nu}
\nonumber\\
&& + \frac{i}{2} \;\tilde \Phi^{\mu(g)} \cdot \tilde \Phi^{(g)}_\mu \;+ \;\tilde {\bar F}^{(h)} \cdot \tilde F^{(h)} \Bigr ].
\end{eqnarray}
It is pretty obvious now to note the following mapping
\begin{eqnarray}
\frac{\partial}{\partial \theta}\; \bigl [\tilde {\cal L}_B\bigr ] = 0 \qquad \Leftrightarrow \qquad
s_{ab} {\cal L}_{\bar B} = - \;\partial_\mu [\bar B \cdot D^\mu \bar C].
\end{eqnarray}
The arguments for the above equivalence between the (4, 2)-dimensional superfield formalism
and the continuous anti-BRST symmetry in the ordinary 4D space is exactly same as we have discussed in the previous section.

Exploiting the following Noether formula for the current in terms
of the generic field $\Psi_i = A_\mu, \Phi_\mu, B_{\mu\nu},
C, \bar C, B, \bar B$ of the Lagrangian density ${\cal L}_{\bar B}$:
\begin{eqnarray}
J^\mu_{(ab)} = (s_{ab} \Psi_i) \; \cdot \Bigl (\frac{\partial {\cal L}_B} {\partial_\mu \Psi_i} \Bigr )
 + \bar B \cdot D^\mu \bar C,
\end{eqnarray}
we obtain the expression for the anti-BRST current $J^\mu_{(ab)}$ as
\begin{eqnarray}
J^\mu_{(ab)} &=& - \bar B \cdot D^\mu \bar C
- \bigl [F^{\mu\nu} - \frac{m}{2}\; \varepsilon^{\mu\nu\eta\kappa} \; 
B_{\eta\kappa} \bigr ] \cdot D_\nu \bar C \nonumber\\
&-& \frac{m}{2}\; \varepsilon^{\mu\nu\eta\kappa}\; (\Phi_\nu \times \bar C) \cdot B_{\eta\kappa}
+ \frac{i}{2}\; \partial^\mu C \cdot (\bar C \times \bar C).
\end{eqnarray}
Using the following Euler-Lagrange equations of motion
\begin{eqnarray}
&& D_\mu \;F^{\mu\nu} + \frac{m}{2}\; \varepsilon^{\nu\mu\eta\kappa}\; \tilde D_\mu\; B_{\eta\kappa}
+ \partial^\nu \bar B = - \;i \;(\bar C \times \partial^\nu C),  \nonumber\\
&& \partial_\mu (D^\mu \bar C) = 0,\; \qquad D_\mu (\partial^\mu C) = 0,\; \qquad F^{\mu\nu} = m \;\varepsilon^{\mu\nu\eta\kappa}\; B_{\eta\kappa},
\nonumber\\
&& \varepsilon^{\mu\nu\eta\kappa}\; \tilde D_\nu\; B_{\eta\kappa} + 2\; m \;\Phi^\mu = 0,\;
\;\; \qquad\;\;  {\cal F}_{\mu\nu} = 0,
\end{eqnarray}
derived from the Lagrangian density ${\cal L}_{\bar B}$, one can, not only prove the conservation law $\partial_\mu J^\mu_{(ab)} = 0$,
but also re-express the above conserved anti-BRST current in a compact form as given below
\begin{eqnarray}
J^\mu_{(ab)} &=& - \bar B \cdot D^\mu \bar C + \partial^\mu \bar B \cdot \bar C + \frac{i}{2} (\partial^\mu C \times \bar C) \cdot \bar C
\nonumber\\
&+& \partial_\nu \;\Bigl [(F^{\nu\mu} + \frac{m}{2}\; \varepsilon^{\mu\nu\eta\kappa} \;B_{\eta\kappa}) \cdot C \Bigr ],
\end{eqnarray}
where we have exploited the Leibnitz rule of the operation of spacetime derivative on a set of combination of local fields.

The above conserved current leads to the definition of the conserved charge. This anti-BRST charge $Q_{ab} = \int d^3 x J^0_{(ab)}$,
is as follows:
\begin{eqnarray}
Q_{ab} = - \;\int \;d^3 x \;\bigl [\bar B \cdot D^0 \bar C - \dot {\bar B} \cdot \bar C
- \frac{i}{2}\; \dot {C} \cdot (\bar C \times \bar C) \bigr ].
\end{eqnarray}
From the definition of the canonical momenta (derived from the Lagrangian density ${\cal L}_{\bar B}$), it can be checked
that the above anti-BRST charge contains the momenta of all the fields except the auxiliary field $B_{\mu\nu}$. To see it
clearly,
we express here the time derivative on $\bar B$ in terms of the other fields, namely;
\begin{eqnarray}
\dot {\bar B} \equiv \partial^0 \bar B = - D_i F^{i0} - \frac{m}{2}\; \varepsilon^{0ijk} \tilde D_i B_{jk}
- i\; (\bar C \times \dot C).
\end{eqnarray}
As a consequence, the conserved and nilpotent anti-BRST charge $Q_{ab}$ generates all the anti-BRST symmetry transformations
for all the fields {\it except} $B_{\mu\nu}$. As argued earlier, the auxiliary field $B_{\mu\nu}$ is completely
different from the Nakanishi-Lautrup type of auxiliary fields $B$ and $\bar B$ in the sense that we obtain the
(anti-)BRST symmetry transformations for the latter fields by the requirements of nilpotency and anticommutativity
of $s_{(a)b}$ but we are unable to do so for the former auxiliary field. This is a novel observation in
our present BRST analysis of the FT model for the non-Abelian TMGT.\\

\noindent
{\large \bf 6. Ghost charge and BRST algebra: a synopsis}\\

\noindent
Let us focus on the ghost part of the coupled (but equivalent) Lagrangian densities. It is clear that
these terms (along with the total Lagrangian densities) are invariant under the following global scale transformations
\begin{eqnarray}
C \to e^{+ \Sigma},\; \qquad \bar C \to e^{- \Sigma},\; \qquad (A_\mu, \Phi_\mu, B_{\mu\nu}) \to 
(A_\mu, \Phi_\mu, B_{\mu\nu}),
\end{eqnarray}
where $\Sigma$ is a global parameter and $\pm$ signs, in the exponentials, denote the ghost number
$(\pm1)$ for $C$ and $\bar C$, respectively.
The above transformations also show that the ghost number for the fields ($A_\mu, \Phi_\mu, B_{\mu\nu}$) is zero. The
infinitesimal version of the above scale transformations leads to the derivation of conserved current
$J^\mu_{(g)}$ and corresponding charge $Q_g = \int d^3 x J^0_{(g)}$ as\footnote{We have taken the Lagrangian
density ${\cal L}_B$ for the computation of the ghost current and corresponding charge. However, one can choose
the other Lagrangian density ${\cal L}_{\bar B}$ as well for this kind of computation to obtain the expressions
for the current and charge.}
\begin{eqnarray}
J^\mu_{(g)} = i \; \bigl [\bar C \cdot D^\mu C - \partial^\mu \bar C \cdot C \bigr ], \qquad
Q_g = i \;\int\; d^3 x\; \bigl [ \bar C \cdot D^0 C - \dot {\bar C} \cdot C \bigr ].
\end{eqnarray}
By exploiting the equations of motion ($\partial_\mu D^\mu C = 0, D_\mu \partial^\mu \bar C = 0$),
derived from the Lagrangian density ${\cal L}_B$, it can be checked that
the above charge and current are conserved and $Q_g$ turns out to be the generator of the infinitesimal version
of the scale transformations listed above (cf. (45)).

We can tap the potential and power of the idea of a generator to derive the BRST algebra amongst
the conserved charges $Q_b, Q_{ab}, Q_g$. For instance, we can check that $s_b \;Q_b = = - i\; \{Q_b, Q_b \} = 0,
s_b\; Q_{ab} = - i \; \{ Q_{ab}, Q_b \} = -i\; (Q_b Q_{ab} + Q_{ab} Q_b) = 0$, etc. Collecting all such
computations, we find that the following standard BRST algebra
\begin{eqnarray}
&& Q_b^2 = 0,\; \qquad\; Q_{ab}^2 = 0,\; \qquad\; 
Q_b Q_{ab} + Q_{ab} Q_b \;\equiv \;\{Q_b, Q_{ab} \} \;= 0, \nonumber\\
&& i\; [Q_g , Q_b] = + Q_b, \quad \;i \;[Q_g, Q_{ab}] = - Q_{ab}, \;\quad i\; [Q_g, Q_b Q_{ab}] = 0,
\end{eqnarray}
emerges from the above conserved charges. It is worth pointing out that

(i) the absolute anticommutativity,
between $Q_b$ and $Q_{ab}$, requires the validity of CF condition, and

(ii) from the above algebra, it
is clear that the ghost numbers for the charges $Q_b, Q_{ab}$ and $Q_b Q_{ab}$ are $+1, - 1$ and $0$, respectively.\\

\noindent
{\large \bf 7. Conclusions}\\

\noindent
We have accomplished our central goal of obtaining the off-shell nilpotent
and absolutely anticommuting (anti-)BRST transformations $s_{(a)b}$
corresponding to the (1-form) YM gauge symmetry transformations (2) for
the original FT Lagrangian density ${\cal L}_0$ of equation (1). In this attempt,
the ``augmented'' BT superfield formalism has played a key role as it has led to the
derivation of CF condition [cf. equation (7)] that is responsible for the
absolute anticommutativity of $s_{(a)b}$ and the derivation of the coupled
(but equivalent) Lagrangian densities [cf. equation (25)]. In this context, a
novel observation is the fact that one is theoretically compelled to invoke GIRs,
in addition to the celebrated HC, in the application of BT superfield formalism
to the description of the 4D topologically massive non-Abelian gauge theory. To
be specific, we obtain the (anti-)BRST transformations for the $B_{\mu\nu}$
and $\Phi_\mu$ fields {\it only} when HC and GIRs blend together in a meaningful manner\footnote{We
christen this modified version of the BT superfield approach
(where HC and GIRs must blend together for proper application) as the ``augmented'' BT superfield formalism.}.

One of the main motivations for our present investigation was to conduct a comparative
study of the FT model within the framework of BRST and superfield formulations
{\it vis-{\`a}-vis} such a kind of study performed for the dynamical non-Abelian
2-form (topologically massive) gauge theory, described by
the following Lagrangian density with the topological ($B \wedge F$) term [6,7,9]
\begin{eqnarray}
{\cal L}_D = - \frac{1}{4} F^{\mu\nu} \cdot F_{\mu\nu} + \frac{1}{12} H^{\mu\nu\eta}
\cdot H_{\mu\nu\eta} + \frac{m}{4} \varepsilon^{\mu\nu\eta\kappa} B_{\mu\nu} \cdot F_{\eta\kappa},
\end{eqnarray}
where the 3-form [$H^{(3)} = \frac{1}{3!} (dx^\mu \wedge dx^\nu\wedge dx^\eta) \; H_{\mu\nu\eta}$]
curvature tensor $H_{\mu\nu\eta}$ is defined in terms of the 1-form gauge field $A_\mu$,
1-form auxiliary field $K_\mu$ and the dynamical 2-form ($B^{(2)} = \frac{1}{2!} (dx^\mu \wedge dx^\nu) B_{\mu\nu}$)
gauge field $B_{\mu\nu}$ as
\begin{eqnarray}
H_{\mu\nu\eta} = D_\mu B_{\nu\eta} + D_\nu B_{\eta\mu} + D_\eta B_{\mu\nu} -
(K_\mu \times F_{\nu\eta} + K_\nu \times F_{\eta\mu} + K_\eta \times F_{\mu\nu}).
\end{eqnarray}
In the above, we have taken the usual definition of the covariant derivative as:
$D_\mu B_{\nu\eta} = \partial_\mu B_{\nu\eta} - (A_\mu \times B_{\nu\eta})$. It was
observed, in the BRST analysis (corresponding to the 1-form YM gauge symmetries)
of this model, that the conserved and nilpotent
(anti-)BRST charges were unable to generate the (anti-)BRST symmetry transformations
for the component $B_{0i}$ of the 2-form dynamical field $B_{\mu\nu}$ as well as the auxiliary
1-form field $K_\mu$, even though, the transformations of these fields were taken into
account in the derivation of the above charges [20]. In contrast, we have shown, in our
present study, that the (anti-)BRST charges of the FT model, are {\it not} capable of
generating the nilpotent (anti-)BRST symmetry transformations corresponding to
the 2-form auxiliary field $B_{\mu\nu}$ of our present non-Abelain version of TMGT.

We find that there is a distinct difference between the Nakanishi-Lautrup type
auxiliary fields $B, \bar B$ and the auxiliary field $B_{\mu\nu}$. As is
evident from our present discussions, the requirements of the off-shell nilpotency
and absolute anticommutativity of the (anti-)BRST symmetry transformations lead
to the determination of these symmetry transformations for $B$ and $\bar B$. On
the contrary, the above requirements are found to be {\it not} good enough to produce
the (anti-)BRST symmetry transformations for the field $B_{\mu\nu}$. The central
reason for this discrepancy is the fact that the momentum, for the field
$B_{\mu\nu}$, does not appear in the expressions for the (anti-)BRST charges,
even though, we perform the BRST analysis in its most general form [cf. equation (24)].
In fact, the gauge-fixing and Faddeev-Popov ghost terms do {\it not} appear for the
fields $B_{\mu\nu}$ and $\Phi_\mu$ in the
most general Lagrangian densities where the basic tenets of BRST formalism have been
exploited in their full generality.

The present model has also been discussed within the framework of the superfield
and BRST formalism in [24] where a tower of auxiliary fields have been invoked
for the consistency of the BRST symmetries, equations of motion and integrability
of the BRST equation. The extended BRST algebra of the FT model has also been derived in [25]
and some of the specific issues, related with this model, have been addressed in it.
However, in these attempts, the CF condition does not appear
which is the root cause for the absolute anticommutativity of the (anti-)BRST symmetry
transformations. One of the key ingredients of the BT superfield approach to BRST formalism
(that is applied to the description of any arbitrary $p$-form gauge theory) is the
very natural derivation of the CF condition. We claim that the appearance of the latter
condition (i.e. CF condition) is the hallmark of any
$p$-form gauge theory described within the framework of BRST formalism.

We have established a deep connection of the CF type restrictions with geometrical objects called gerbes
in a couple of papers where we have discussed the Abelian 2-form and 3-form gauge theories
[16,17]. We plan to establish connections of the CF type restrictions, appearing in the
discussion of the non-Abelian TMGTs, with the concept of  gerbes. Furthermore,
it would be nice endeavor to exploit the tensor gauge symmetries of our present model
within the framework of BRST and superfield formalisms. A universal Lagrangian density for
the massive Yang-Mills theories has been proposed in [26] which does not take recourse
to the Higgs mechanism for the generation of mass. It would be interesting to study
this Lagrangian density within our superfield and BRST formalisms which crucially depend
on the ``augmented'' version of the BT superfield formalism applied to any arbitrary
$p$-form gauge theory. These are some of
the issues that are being pursued at the moment and our results would be reported
in our future publications [27].\\

\noindent
{\large \bf Acknowledgements}\\

\noindent
A major part of this work was completed at the AS-ICTP, Trieste, Italy, during the summer visit.
It is our pleasure and honor
to thank the Director, AS-ICTP, Trieste, for the invitation and
warm hospitality at the AS-ICTP. Fruitful conversations with L. Bonora (SISSA) and K. S.
Narain (AS-ICTP) are thankfully acknowledged, too.

\end{document}